
\normalbaselineskip = 12 pt
\magnification 1200
\hsize = 15 truecm \vsize = 22 truecm \hoffset = 1.0 truecm
\font\petit=cmr8
\def\half{{1\over2}}
\def\d{\partial}
\def\<{\langle}
\def\>{\rangle}
\rightline{LANDAU-94-TMP-4}
\rightline{hep-th/9406118}
\rightline{June 1994}
\vskip 4 truecm
\centerline{\bf SECTORS OF MUTUALLY LOCAL FIELDS}
\centerline{\bf IN INTEGRABLE MODELS}
\centerline{{\bf OF QUANTUM FIELD THEORY}%
\footnote{$^*$}%
{The work was supported in part by Landau Institute Foundation,
International Science Foundation, and Russian Foundation for
Fundamental In\-ves\-ti\-ga\-tions.}
}
\bigskip
\centerline{M. Yu. Lashkevich%
\footnote{$^+$}{E-mail: lashkevi@cpd.landau.free.net}}
\centerline{\it Landau Institute for Theoretical Physics,}
\centerline{\it Kosygina 2, GSP-1, 117940 Moscow V-334, Russia}
\bigskip
{\petit It is known that 2D field theories admit several sectors
of mutually local fields so as two fields from different sectors
are generally mutually nonlocal. We show that any one-particle
integrable model with ${\bf Z_2}$ symmetry contains three sectors:
bosonic, fermionic
and `disorder' one. We generalize the form factor axioms to fermionic
and `disorder' sectors. For the particular case of the
sinh-Gordon model we obtain several form factors in these sectors.}
\vfill\eject
\noindent
{\bf 1. Introduction}
\medskip\noindent
Any quantum field theory consists of two structures: quantum
mechanics (state space and Hamiltonian) and a set of mutually
local fields. Any two-dimensional field theory with given
quantum-mechanical structure admits different local field
realizations.$^1$ The most wellknown examples are the following.
The Ising model can be performed either as a free fermion
or as an interacting boson.$^{2,3}$ The symplectic
model can be repesented either as a free neutral boson or as
an interacting fermion.$^3$ The sin-Gordon model and the Thirring
model coincide quantum-mechanically,$^4$ but their local field contents
are different: the bosonic kink fields create the same asymptotic
states as the Thirring fermions.$^5$

Recall some properties of the Ising and symplectic models.$^3$
The Ising model in the scaling limit above the critical point
can be considered as
a free Majorana fermion $\psi(x)$ with the action
$$
{\cal A}=\half\int d^2x\,\left(i\overline{\psi}\gamma^\mu
\d_\mu\psi-M\overline{\psi}\psi\right).
$$
On the other hand there are two bosonic fields, $\varphi(x)$ (order
parameter) and $\mu(x)$ (disorder parameter), connected with the
fermion $\psi(x)$ nonlinearly and nonlocally. Their simulaneous
commutation relations are
$$
\eqalign{
\varphi(x)\psi(y)
&=\epsilon(x^1-y^1)\psi(y)\varphi(x),
\cr
\mu(x)\psi(y)
&=-\epsilon(x^1-y^1)\psi(y)\mu(x),
\cr
\mu(x)\varphi(y)
&=-\epsilon(x^1-y^1)\varphi(y)\mu(x)\quad\hbox{for}\quad x^1\neq y^1,
}\eqno(1.1)
$$
where $(x^0,x^1)$ are Minkowski coordinates and
$\epsilon(\xi)\equiv\hbox{sign }\xi$. This means that these
fields are mutually nonlocal. It is important to note that
the fermion alone creates the whole state space, and the
order parameter acts in it, but does not create
any new space. The order parameter $\varphi(x)$ creates the same
asymptotic states as fermions $\psi(x)$, but the $S$-matrix of the
bosonic field is nontrivial. The $S$-matrix of $N$ bosons is
equal to $(-1)^{\half N(N-1)}$. The disorder parameter $\mu(x)$
does not create any asymptotic states.

So there are at least three sectors in the Ising model:
`bosonic' sector containing fields commuting with the order
parameter, `fermionic' sector containing fields commuting
or anticommuting with the fermion, and `disorder' sector of
fields commuting with the disorder parameter.

Similar situation is observed in the symplectic model. Bosonic
sector contains a free neutral boson with the action
$$
{\cal A}=\half\int d^2x\,\left(\d_\mu\varphi\,\d^\mu\varphi
-M^2\varphi^2\right).
$$
Fermionic and `disorder' sectors contain interacting fermion $\psi(x)$
and bosonic `disorder' parameter $\mu(x)$ respectively. The $S$-matrix
of $N$ fermions is equal to $(-1)^{\half N(N-1)}$. The fields
$\psi(x)$, $\varphi(x)$ and $\mu(x)$ possess the same simultaneous
commutation relations (1.1).

On the basis of these two examples and Ref. 1 we can formulate
the

{\bf Conjecture.} {\it Any integrable 2D field theory with
${\bf Z_2}$ symmetry of changing the sign at
the `elementary' field $\varphi(x)$
($\varphi(x)\longrightarrow-\varphi(x)$)
contains at least three sectors of local fields.
Two `dual' sectors contain $n$-component fields $\varphi(x)$
and $\psi(x)$ respectively, and a `disorder' sector
contains an $n$-component bosonic field $\mu(x)$.
The following conditions hold:

1) the fields $\varphi_i$, $\psi_j$ and $\mu_k$ are pairwise mutually
nonlocal;

2) if $\varphi_i(x)$ is a boson, then $\psi_i(x)$ is a fermion and
the commutation relations (1.2) hold for $\varphi_i$, $\psi_i$
and $\mu_i$; if $\varphi_i(x)$ is a fermion, $\psi_i(x)$ is a boson;

3) the fields $\varphi(x)$ and $\psi(x)$ create the same assymptotic
states, their $S$-matrix is factorizable and corresponding
two-particle $S$-matrices $S_\varphi(\theta)$ and $S_\psi(\theta)$
($\theta$ is the difference of rapidities) are connected:
$$
S_\psi(\theta)=-S_\varphi(\theta);
\eqno(1.2)
$$

3) the `disorder' field does not produce asymptotic states.}

In this paper we substantiate this conjecture in the simplest
case of the one-component interacting fields $\varphi(x)$ and $\psi(x)$,
and illustrate it by the sinh-Gordon model.

In Sec. 2 we recall some fundamental facts about form factors.
We use the formulation through the Zamolodchikov algebra,$^6$ because
it clarifies some aspects of the form factor axioms. In Sec. 3 we formulate
Karowski--Weisz--Smirnov axioms for form factors$^{7,8}$ in slightly
generalized form and obtain the sectors of mutually local fields.
Sec. 4 reviews trivial examples: the Ising and the symplectic models.
A more complicated example --- sinh-Gordon model --- is
analysed in Sec. 5. Conclusions and unsolved problems are listed
in Sec. 6.
\medskip\noindent
{\bf 2. Zamolodchikov algebra and form factors}

\nobreak\medskip\nobreak\noindent
Let $S(\theta)$ be an analitical single-valued function
without poles in the strip $0<\hbox{Im }\theta\leq\pi$ and
satisfying the equations
$$
S(-\theta)=S(\theta+i\pi)=S^{-1}(\theta).
\eqno(2.1)
$$
It is necessary to stress that we do not identify the function
$S(\theta)$ with any two-particle $S$-matrix. The connection
between $S(\theta)$ and $S$-matrix will be established in the
following section.

The Zamolodchikov algebra$^6$ is generated by symbols $V(\theta)$,
$V^+(\theta)$ ($\theta\in{\bf R}$) satisfying the equations
$$
\eqalign{
V(\theta_1)V(\theta_2)
&=S(\theta_1-\theta_2)V(\theta_2)V(\theta_1),
\cr
V^+(\theta_1)V^+(\theta_2)
&=S(\theta_1-\theta_2)V^+(\theta_2)V^+(\theta_1),
\cr
V(\theta_1)V^+(\theta_2)
&=S^{-1}(\theta_1-\theta_2)V^+(\theta_2)V(\theta_1)
+2\pi\delta(\theta_1-\theta_2).
}\eqno(2.2)
$$
We can introduce normal ordering, $:\cdots:$, by the rules
$$
\eqalign{
&:V^+(\theta_1)\cdots V^+(\theta_m)V(\vartheta_1)\cdots V(\vartheta_n):
=V^+(\theta_1)\cdots V^+(\theta_m)V(\vartheta_1)\cdots V(\vartheta_n),
\cr
&:\cdots V(\theta_1)V^+(\theta_2)\cdots:
=S^{-1}(\theta_1-\theta_2):\cdots V^+(\theta_2)V(\theta_1)\cdots:.
}\eqno(2.3)
$$
Now we shall describe a quantum-mechanical system. The vacuum
$|0\>$ is defined by the equation
$$
V(\theta)|0\>=0,\quad\theta\in{\bf R}.
\eqno(2.4)
$$
The state space is spanned on the basis
$$
V^+(\theta_1)\cdots V^+(\theta_n)|0\>,
\quad n=0,1,2,\cdots,\quad\theta_i\in{\bf R}.
\eqno(2.5)
$$
Evidently
$$
\eqalign{
\<V(\theta_1)V(\theta_2)\>
&=\<V^+(\theta_1)V^+(\theta_2)\>=\<V^+(\theta_1)V(\theta_2)\>=0,
\cr
\<V(\theta_1)V^+(\theta_2)\>
&=2\pi\delta(\theta_1-\theta_2)
}\eqno(2.6)
$$
($\<\cdots\>\equiv\<0|\cdots|0\>$), and the Wick theorem holds.

The Hamiltonian is given by
$$
{\cal H}=\int_{-\infty}^\infty{d\theta\over2\pi}\,P_0(\theta)
V^+(\theta)V(\theta),
\eqno(2.7)
$$
where $P_0(\theta)$ is the time component of the covector
$$
P(\theta)=(M\cosh\theta,-M\sinh\theta),
\eqno(2.8)
$$
$M$ is a parameter. Also we shall use the designation
$$
P(\theta_1,\cdots,\theta_n)=M\sum_{i=1}^n(\cosh\theta_i,-\sinh\theta_i).
\eqno(2.9)
$$
The system described above is linear and, therefore, its solution
is evident. The only problem is to construct local fields
in this theory. If $S(\theta)\equiv1$ (boson) or $S(\theta)\equiv-1$
(fermion), it is easy to introduce local fields
$$
\varphi(x)=\int_C{d\theta\over2\pi}e^{-iP(\theta)x}V(\theta)
\quad\hbox{for}\quad S(\theta)=1
\eqno(2.10)
$$
and
$$
\psi_\pm(x)=\sqrt{M}\int_C{d\theta\over2\pi}e^{\mp\half({i\pi\over2}+\theta)}
e^{-iP(\theta)x}V(\theta)
\quad\hbox{for}\quad S(\theta)=-1.
\eqno(2.11)
$$
The contour $C$ is a sum of two straight lines:
$C=(-\infty-i0,+\infty-i0)+(-\infty-i\pi+i0,+\infty-i\pi+i0)$;
the lines are directed to the right.
The sense of the infinitesimal shifts will be clarified later.
Here we used the following designation
$$
V(\theta-i\pi)=V^+(\theta),\quad\theta\in{\bf R}.
\eqno(2.12)
$$
The fields $\varphi(x)$ and $\psi_\pm(x)$ are usual bosonic
and fermionic local fields of mass $M$.
For general $S(\theta)$ we consider a field
$$
\phi(x)=\sum_{n=0}^\infty{1\over n!}\int_C{d\theta_1\over2\pi}
\cdots\int_C{d\theta_n\over2\pi}e^{-iP(\theta_1,\cdots,\theta_n)x}
F_\phi^{(n)}(\theta_1,\cdots,\theta_n):V(\theta_n)\cdots V(\theta_1):.
\eqno(2.13)
$$
Here $F_\phi^{(n)}(\theta_1,\cdots,\theta_n)$ are some analitical
single-valued functions satisfying the equation
$$
F_\phi^{(n)}(\cdots,\theta_i,\theta_{i+1},\cdots)
=S(\theta_i-\theta_{i+1})F_\phi^{(n)}(\cdots,\theta_{i+1},\theta_i,\cdots).
\eqno(2.14)
$$
Eq. (2.13) gives nearly general form of a Heisenberg operator
$$
i{\d\phi(x)\over\d x^0}=[\phi(x),{\cal H}\,].
\eqno(2.15)
$$
The operator $\phi(x)$ is Hermitain if and only if
$$
F_\phi^{(n)}(\theta_1,\cdots,\theta_n)
=F_\phi^{(n)*}(\theta_n+i\pi,\cdots,\theta_1+i\pi)
\hbox{ for }\theta_1,\cdots,\theta_n\in{\bf R}\cup({\bf R}-i\pi).
\eqno(2.16)
$$

Functions $F_\phi^{(n)}(\theta_1,\cdots,\theta_n)$ are referred to
as form factors of the operator $\phi(x)$. It is easy to represent
correlation functions in form of infinite series using the Wick
theorem. For example,
$$
\eqalign{
\<\phi_1(x)\phi_2(y)\>
&=\sum_{n=0}^\infty{1\over n!}
\int_{-\infty}^\infty{d\theta_1\over2\pi}\cdots
\int_{-\infty}^\infty{d\theta_n\over2\pi}
e^{-iP(\theta_1,\cdots,\theta_n)(x-y)}
\cr
&\times F_{\phi_1}^{(n)}(\theta_1,\cdots,\theta_n)
F_{\phi_2}^{(n)}(\theta_n-i\pi,\cdots,\theta_1-i\pi).
}$$
There are two important problems to be solved. Firstly, what is the
condition of mutual locality of two fields? Secondly, what asymtotical
states do the fields (2.13) create? The Karowski--Weicz--Smirnov
axioms answer these questions.
\medskip\noindent
{\bf 3. Form factor axioms and sectors of local fields}

\nobreak\medskip\nobreak\noindent
In this section we consider Karowski--Weisz--Smirnov axioms$^{7,8}$
as conditions of locality. We follow Smirnov,$^8$ but slightly
weaken the axioms. It will allow us to consider fermionic
and `disorder' fields.

Namely, let us impose the following conditions on form factors:

0. Definite Lorentz spin $s_\phi$
$$
F_\phi^{(n)}(\theta_1+\vartheta,\cdots,\theta_n+\vartheta)
=e^{-s_\phi\vartheta}F_\phi^{(n)}(\theta_1,\cdots,\theta_n).
\eqno(3.1)
$$

1. Consistency with the Zamolodchikov algebra (Eq. (2.14)):
$$
F_\phi^{(n)}(\cdots,\theta_i,\theta_{i+1},\cdots)
=S(\theta_i-\theta_{i+1})F_\phi^{(n)}(\cdots,\theta_{i+1},\theta_i,\cdots).
\eqno(3.2)
$$

2. Cyclic eqation
$$
F_\phi^{(n)}(\theta_1+2\pi i,\theta_2,\cdots,\theta_n)
=\lambda_\phi F_\phi^{(n)}(\theta_2,\cdots,\theta_n,\theta_1),
\quad\lambda_\phi=\pm1.
\eqno(3.3)
$$

3. The only poles in the strip $0\leq\hbox{Im }(\theta_i-\theta_j)
\leq\pi$, $i<j$ are $\theta_i=\theta_j+i\pi$ and
$$
\eqalign{
F_\phi^{(n+2)}(\vartheta+i\pi+\varepsilon,\vartheta,
\theta_1,\cdots,\theta_n)
&\simeq{i\over\varepsilon}
\left[1-\zeta_\phi\prod_{i=1}^nS(\vartheta-\theta_i)\right]
F_\phi^{(n)}(\theta_1,\cdots,\theta_n),
\cr
\varepsilon\longrightarrow0,
&\quad\zeta_\phi=\pm1.
}\eqno(3.4)
$$
Later we shall see that
$$
\lambda_\phi=\zeta_\phi,
\eqno(3.3')
$$
but we shall ignore it for a while.

Notice that (in absence of bound states) the axioms do not
connect the form factors $F_\phi^{(2n)}$ and $F_\phi^{(2n-1)}$.
Therefore, it is sufficient to  consider such fields $\phi$ that
contain nonzero form factors of either even or odd degree only.
It is natural to call these fields  even or odd respectively.

First of all let us calculate the physical $S$-matrices.
It can be shown that even fields do not create asymptotic
states. Consider $t\longrightarrow\pm\infty$ asymptotics
of odd fields:$^{3,8}$
$$
\eqalign{
\phi_{out/in}(\theta)
&=\lim_{x^0\rightarrow\pm\infty}{i\epsilon(e^\theta)\over2}
\int_{-\infty}^\infty dx^1\left(e^{iP(\theta)x}\d_0\phi(x)
-\phi(x)\,\d_0e^{iP(\theta)x}\right)
\cr
&=\lim_{t\rightarrow\pm\infty}{\epsilon(e^\theta)\over2}
\sum_{n=0}^\infty{1\over(2n+1)!}\int_C{d\theta_1\over2\pi}
\cdots\int_C{d\theta_{2n+1}\over2\pi}
\cr
&\times F_\phi^{(2n+1)}(\theta_1,\cdots,\theta_{2n+1})
:V(\theta_{2n+1})\cdots V(\theta_1):
\cr
&\times\left[\sum_i\left(e^{\theta_i}+e^{-\theta_i}\right)
+e^\theta+e^{-\theta}\right]
\cdot2\pi\delta\left[e^\theta-e^{-\theta}
-\sum_i\left(e^{\theta_i}-e^{-\theta_i}\right)\right]
\cr
&\times\exp\left\{iM{t\over2}\left[e^\theta+e^{-\theta}
-\sum_i\left(e^{\theta_i}+e^{-\theta_i}\right)\right]\right\}.
}\eqno(3.5)
$$
Only the poles contribute to asymptotic fields according to
the formula
$$
\lim_{t\rightarrow\pm\infty}{-i\over x-i0}e^{-ixat}
=\Theta(\mp a)\cdot2\pi\delta(x),
\eqno(3.6)
$$
with $\Theta(x)$ being the Heaviside theta-function.
Consider the pole $\theta_{2i-1}\longrightarrow\theta_{2i}+i\pi$,
$i=1,2,\cdots,n$. Eq. (3.4) leads to
$$
F_\phi^{(2n+1)}(\theta_1,\cdots,\theta_{2n+1})
\simeq i^nF_\phi^{(1)}(\theta_{2n+1})
\prod_{i=1}^n{
1-\zeta_\phi S(\theta_{2i-1}-\theta_{2n+1}
\over\theta_{2i-1}-\theta_{2i}-i\pi}.
\eqno(3.7)
$$
Because of the $\delta$-function in Eq. (3.5)
$$
e^\theta-e^{\theta_{2n+1}}=\sum_{i=1}^n
\left(e^{\theta_{2i-1}}+e^{\theta_{2i}}\right)
{1-e^{-\theta_{2i-1}-\theta_{2i}}\over1-e^{-\theta-\theta_{2i+1}}}.
\eqno(3.8)
$$
Substituting Eq. (3.8) into the exponential in Eq. (3.5), and
Eq. (3.7) into Eq. (3.5), using Eq. (3.6) [infenitesimal shifts
in the contour $C$ produce $-i0$ for Eq. (3.6)], and taking
into account a combinatorial factor $(2n+1)!/n!$
from permutations of $\theta_i$ in Eq. (3.7), we get
asymptotic fields
$$
\phi_{out/in}(\theta)
=F_\phi^{(1)}(\theta):V(\theta)
e^{\int_{-\infty}^\infty{d\theta'\over2\pi}
\Theta(\pm\ {\rm Re}\ (\theta-\theta'))
\left[\zeta_\phi S(\theta-\theta')-1\right]V^+(\theta')V(\theta')}:.
\eqno(3.9)
$$
Consider asymptotic states:
$$
|\theta\>_{out/in}=\phi_{out/in}^+(\theta)|0\>
=F_\phi^{(1)}(\theta)V^+(\theta)|0\>,
$$
because only the 0th term in the decomposition of the exponential
acts on the vacuum nontrivially. Similarly, for $\theta_1>\theta_2$
$$
\eqalign{
|\theta_1,\theta_2\>_{out}
&=\zeta_\phi S(\theta_1-\theta_2)
F_\phi^{(1)}(\theta_1)F_\phi^{(1)}(\theta_2)
V^+(\theta_1)V^+(\theta_2)|0\>,
\cr
|\theta_1,\theta_2\>_{in}
&=F_\phi^{(1)}(\theta_1)F_\phi^{(1)}(\theta_2)
V^+(\theta_1)V^+(\theta_2)|0\>,
}$$
and we see that the physical $S$-matrix is given by
$$
S_\phi(\theta)=\zeta_\phi S(\theta).
\eqno(3.10)
$$
It is easy to cheque that Eq. (3.10) holds also for $\theta_1<\theta_2$,
and that for $n$ particles the physical $S$-matrix is
$$
S_{\phi,{\rm phys}}^{(n)}(\theta_1,\cdots,\theta_n)
=\prod_{i<j}S_\phi(\theta_i-\theta_j).
\eqno(3.11)
$$
It may seem strange that the $S$-matrix depends on the field
$\phi(x)$ while the asymptotic states are independent of it
up to normalization factors. However, the $S$-matrix
depends on the choice of the `unperturbed' Hamiltonian (or,
in other words, of the interaction representation). Fixing
a field $\phi$ we fix the `unperturbed' Hamiltonian with
respect to which the field $\phi$ is free.

Now we shall cheque locality.$^8$ Let $x^0=y^0$, $x^1>y^1$.
Then in an integral $\int_{-\infty}^\infty{d\xi\over2\pi}
e^{-iP(\xi)(x-y)}(\cdots)$ one can shift the contour upward
till $i\pi$ preserving convergence of the integral. Consider
the product
$$
\eqalign{
\phi_1(x)\phi_2(y)
&=\sum_{m,n,k=0}^\infty{1\over m!n!k!}
\int_{-\infty}^\infty{d^k\xi\over(2\pi)^k}
\int_C{d^m\theta\over(2\pi)^m}\int_C{d^n\vartheta\over(2\pi)^n}
\cr
&\times\exp\left[-iP(\xi)(x-y)-iP(\theta)x-iP(\vartheta)y\right]
\cr
&\times F_{\phi_1}^{(m+k)}(\xi_1-i0,\cdots,\xi_k-i0,\theta_1,\cdots\theta_m)
\cr
&\times F_{\phi_2}^{(n+k)}(\vartheta_n,\cdots,\vartheta_1,
\xi_k-i\pi+i0,\cdots,\xi_1-i\pi+i0)
\cr
&\times:V(\theta_m)\cdots V(\theta_1)V(\vartheta_1)\cdots V(\vartheta_n):.
}$$
We used the usual rule for calculating the normal form of
two normally ordering multipliers. Now we shift the contours
over $\xi_i$ up: $\int_{-\infty}^\infty{d^k\xi\over(2\pi)^k}
\longrightarrow\int_{-\infty+i\pi}^{\infty+i\pi}{d^k\xi\over(2\pi)^k}$.
The contour catch on the poles (3.4). Calculating residues of
the poles and regrouping the terms we get
$$
\eqalign{
&\phi_1(x)\phi_2(y)=\sum_{m,n,k=0}^\infty{1\over m!n!k!}
\int_{-\infty}^\infty{d^k\xi\over(2\pi)^k}
\int_C{d^m\theta\over(2\pi)^m}\int_C{d^n\vartheta\over(2\pi)^n}
\cr
&\times\exp\left[iP(\xi)(x-y)-iP(\theta)x-iP(\vartheta)y\right]
\cr
&\times F_{\phi_1}^{(m+k)}(\xi_1+i\pi+i0,\cdots,
\xi_k+i\pi+i0,\theta_1,\cdots\theta_m)
\cr
&\times F_{\phi_2}^{(n+k)}(\vartheta_n,\cdots,\vartheta_1,
\xi_k-i0,\cdots,\xi_1-i0)
\cr
&\times\sum_{l=1}^n(-)^{l-1}\sum_{j_1<\cdots<j_l}^n
\prod_{r=1}^l\left[1-\zeta_\phi
\prod_{i=1}^kS(\vartheta_{j_r}-\xi_i-i\pi)\prod_{i=1}^m
S(\vartheta_{j_r}-\theta_i)
\right]
\cr
&\times:V(\theta_m)\cdots V(\theta_1)V(\vartheta_1)\cdots V(\vartheta_n):.
}$$
The double sum here can be calculated:
$$
\sum_{l=1}^n(-)^{l-1}\sum_{j_1<\cdots<j_l}^n\prod_{l=1}^l
(1-\alpha_{j_r})=\prod_{j=1}^n\alpha_j.
$$
Doing cyclic transposition in $F_{\phi_2}^{(n+k)}$ according to
Eq. (3.3), pulling $\xi$'s through $\theta$'s according to
Eq. (3.2) in $F_{\phi_1}^{(m+k)}$, and reordering Zamolodchikov
operators, we obtain
$$
\eqalign{
\phi_1(x)\phi_2(y)
&=\sum_{m,n,k=0}^\infty{\lambda_{\phi_1}^k\zeta_{\phi_1}^n\over m!n!k!}
\int_{-\infty}^\infty{d^k\xi\over(2\pi)^k}
\int_C{d^m\theta\over(2\pi)^m}\int_C{d^n\vartheta\over(2\pi)^n}
\cr
&\times\exp\left[-iP(\xi)(y-x)-iP(\theta)x-iP(\vartheta)y\right]
\cr
&\times F_{\phi_2}^{(n+k)}(\xi_1-i0,\cdots,
\xi_k-i0,\vartheta_1,\cdots,\vartheta_n)
\cr
&\times F_{\phi_1}^{(m+k)}(\theta_m,\cdots,\theta_1,
\xi_k-i\pi+i0,\cdots,\xi_1-i\pi+i0)
\cr
&\times:V(\vartheta_n)\cdots V(\vartheta_1)V(\theta_1)\cdots V(\theta_m):.
}\eqno(3.12)
$$
We see that
$$
\eqalignno{
\phi_1(x)\phi_2(y)
&=\phi_2(y)\phi_1(x),\quad x^0=y^0,\ x^1>y^1,
\quad\hbox{if }\lambda_{\phi_1}^k\zeta_{\phi_1}^n=1,
&(3.13a)
\cr
\phi_1(x)\phi_2(y)
&=-\phi_2(y)\phi_1(x),\quad x^0=y^0,\ x^1>y^1,
\quad\hbox{if }\lambda_{\phi_1}^k\zeta_{\phi_1}^n=-1.
&(3.13b)}
$$
For definite commutation relations
one of the conditions must hold for all nonzero terms in Eq. (3.12).
Because $k$ is an arbitrary number and $(-1)^{k+n}$ is fixed by
the evenness of the field $\phi_2(x)$, we get
$$
\lambda_\phi=\zeta_\phi.
\eqno(3.3')
$$

It is easy to cheque from consistency of the axioms [and Eq. $(3.3')$]
that $s_\phi\in{\bf Z}$ for an even field $\phi(x)$ and
$$
\zeta_\phi=(-1)^{2s_\phi}\quad\hbox{(odd)}
\eqno(3.14)
$$
for odd one.

If $\phi_2(x)$ is even, the condition (3.13a) holds.
If $\phi_2(x)$ is odd, the condition (3.13a) holds for $\zeta_{\phi_1}=1$,
and the condition (3.13b) holds for $\zeta_{\phi_1}=-1$. We
reproduce the spin-statistics correspondence.%
\footnote{$^a$}{To avoid confusion, note that `statistics' only
means here the commutation relations, but not the statistics
of the gas which is governed by $S(0)=\pm1$.}

Let $\tau(x)$ be an even field with $\zeta_\tau=1$,
$\varphi(x)$ an odd field with $\zeta_\varphi=1$, $\mu(x)$ an even field
with $\zeta_\mu=-1$, and $\psi(x)$ an odd field with $\zeta_\psi=-1$.
We get the simulaneous commutation relations
$$
\eqalign{
\tau(x)\tau(y)&=\tau(y)\tau(x),
\cr
\tau(x)\varphi(y)&=\varphi(y)\tau(x),
\cr
\tau(x)\mu(y)&=\mu(y)\tau(x),
\cr
\tau(x)\psi(y)&=\psi(y)\tau(x),
\cr
\varphi(x)\varphi(y)&=\varphi(y)\varphi(x),
\cr
\varphi(x)\mu(y)&=\epsilon(x^1-y^1)\mu(y)\varphi(x),
\cr
\varphi(x)\psi(y)&=\epsilon(x^1-y^1)\psi(y)\varphi(x),
\cr
\mu(x)\mu(y)&=\mu(y)\mu(x),
\cr
\psi(x)\mu(y)&=\epsilon(x^1-y^1)\mu(y)\psi(x),
\cr
\psi(x)\psi(y)&=-\psi(y)\psi(x).
}\eqno(3.15)
$$
Let $[\tau]$, $[\varphi]$, $[\mu]$, $[\psi]$ be spaces of fields
with the same evenness and $\zeta_\phi$ as $\tau$, $\varphi$, $\mu$, $\psi$
respectively. From Eq. (3.15) we obtain three sectors of
mutually local fields:

`bosonic' sector $\{[\tau],[\phi]\}$;

`fermionic' sector $\{[\tau],[\psi]\}$;

`disorder' sector $\{[\tau],[\mu]\}$.

Now we shall consider examples.
\medskip\noindent
{\bf 4. Simple examples: Ising and symplectic models}

\nobreak\medskip\nobreak\noindent
Here we recall some formulas concerning the Ising and symplectic models$^3$
and cheque that their form factors satisfy Eqs. (3.1)--(3.4).

We begin with the Ising model. The Zamolodchikov algebra of this
model is the Klifford one [$S(\theta)=-1$]. Local fermion,
disorder and spin operators are given by
$$
\eqalign{
\psi_\pm(x)
&=\sqrt{M}\int_C{d\theta\over2\pi}e^{\mp\half(\theta+{i\pi\over2})}
e^{-iP(\theta)x}V(\theta),
\cr
\mu(x)
&=:\exp\half\rho_F(x):,
\cr
\varphi(x)
&=:\psi_0(x)\exp\half\rho_F(x):,
}$$
where
$$
\eqalign{
\rho_F(x)
&=-i\int_C{d\theta_1\over2\pi}\int_C{d\theta_2\over2\pi}
\tanh\half(\theta_1-\theta_2)\cdot e^{-iP(\theta_1,\theta_2)x}
V(\theta_1)V(\theta_2),
\cr
\psi_0(x)
&=\int_C{d\theta\over2\pi}e^{-iP(\theta)x}V(\theta).
}$$
It means that a unique nonzero form factor of the fermion
is given by
$$
F_{\psi_\pm}^{(1)}(\theta)=\sqrt{M}e^{\mp\half(\theta+{i\pi\over2})},
\quad s_{\psi_\pm}=\pm\half.
\eqno(4.1)
$$
Nonzero form factors of the disorder operators are
$$
\eqalign{
F_\mu^{(2n)}(\theta_1,\cdots,\theta_{2n})
&=(-)^n{(2n)!\over n!}\hbox{ Alt }
\prod_{i=1}^n\tanh\half(\theta_{2i-1}-\theta_{2i})
\cr
&=(-)^n\prod_{i<j}^{2n}\tanh\half(\theta_i-\theta_j),
\quad n=0,1,2,\cdots,
}\eqno(4.2)
$$
where Alt means antisymmetrization in $\theta$'s.
Nonzero form factors of the spin operator are given by$^{3,9}$
$$
\eqalign{
F_\varphi^{(2n+1)}(\theta_1,\cdots,\theta_{2n+1})
&=(-)^n{(2n+1)!\over n!}\hbox{ Alt }
\prod_{i=1}^n\tanh\half(\theta_{2i}-\theta_{2i+1})
\cr
&=(-)^n\prod_{i<j}^{2n+1}\tanh\half(\theta_i-\theta_j),
\quad n=0,1,2,\cdots,
}\eqno(4.3)
$$

The energy-momentum tensor gives an example of a field from $[\tau]$.
In light-cone coordinates $x^\pm=\half(x^0\pm x^1)$ it is given by
$$
\eqalign{
T_{+-}(x)
&=iM\psi_+(x)\psi_-(x),
\cr
T_{++}(x)
&=-i:\psi_+(x)\d_+\psi_+(x):,
\cr
T_{--}(x)
&=-i:\psi_-(x)\d_-\psi_-(x):.
}$$
Its nonzero form factors are
$$
\eqalign{
F_{T_{+-}}^{(2)}(\theta_1,\theta_2)
&=iM^2\sinh\half(\theta_1-\theta_2),
\cr
F_{T_{++}}^{(2)}(\theta_1,\theta_2)
&=-iM^2e^{-\theta_1-\theta_2}\sinh\half(\theta_1-\theta_2),
\cr
F_{T_{--}}^{(2)}(\theta_1,\theta_2)
&=-iM^2e^{\theta_1+\theta_2}\sinh\half(\theta_1-\theta_2).
}\eqno(4.4)
$$

For the symplectic model the Zamolodchikov algebra coincides
with the Hei\-sen\-berg algebra [$S(\theta)=1$]. Boson, disorder
and fermionic order fields are given by
$$
\eqalign{
\varphi(x)
&=\int_C{d\theta\over2\pi}e^{-iP(\theta)x}V(\theta),
\cr
\mu(x)
&=:\exp\half\rho_B(x):,
\cr
\psi_\pm(x)
&=:\varphi_\pm(x)\exp\half\rho_B(x):,
}$$
where
$$
\eqalign{
\rho_B(x)
&=-2\int_C{d\theta_1\over2\pi}\int_C{d\theta_2\over2\pi}
{e^{-iP(\theta_1,\theta_2)x}\over\cosh\half(\theta_1-\theta_2)}
V(\theta_1)V(\theta_2),
\cr
\varphi_\pm(x)
&=\sqrt{M}\int_C{d\theta\over2\pi}e^{\mp\half(\theta+{i\pi\over2})}
e^{-iP(\theta)x}V(\theta).
}$$
Corresponding nonzero form factors are
$$
\eqalign{
F_\varphi^{(1)}(\theta)
&=1,
\cr
F_\mu^{(2n)}(\theta_1,\cdots,\theta_{2n})
&={(-2)^n(2n)!\over n!}\hbox{ Sym }
\prod_{i=1}^n\cosh^{-1}\half(\theta_{2i-1}-\theta_{2i}),
\cr
F_{\psi_\pm}^{(2n+1)}(\theta_1,\cdots,\theta_{2n+1})
&\cr
=\sqrt{M}
&{(-2)^n(2n+1)!\over n!}\hbox{ Sym }
e^{\mp\half(\theta_1+{i\pi\over2})}
\prod_{i=1}^n\cosh^{-1}\half(\theta_{2i}-\theta_{2i+1}),
}\eqno(4.5)
$$
where Sym means symmetrization in $\theta$'s.

The energy-momentum tensor
$$
\eqalign{
T_{+-}(x)
&=M^2:\varphi^2:(x),
\cr
T_{++}(x)
&=:(\d_+\varphi)^2:(x),
\cr
T_{--}(x)
&=:(\d_-\varphi)^2:(x)
}$$
belongs to the space $[\tau]$ and has the form factors
$$
\eqalign{
F_{T_{+-}}^{(2)}(\theta_1,\theta_2)
&=M^2,
\cr
F_{T_{++}}^{(2)}(\theta_1,\theta_2)
&=-M^2e^{-\theta_1-\theta_2},
\cr
F_{T_{--}}^{(2)}(\theta_1,\theta_2)
&=-M^2e^{\theta_1+\theta_2},
}\eqno(4.6)
$$
It is easy to cheque directly that all form factors (4.1)--(4.6)
satisfy the generalized form factor axioms.
\medskip\noindent
{\bf 5. sinh-Gordon model}

\nobreak\medskip\nobreak\noindent
Consider a more complicated example: the sinh-Gordon model with the
action
$$
{\cal A}=\int d^2x\left(\half\d_\mu\varphi\d^\mu\varphi
-{\alpha\over\beta^2}\cosh\beta\varphi\right).
\eqno(5.1)
$$
The parameter $\alpha$ depends on cutoff and is not
essential. Really the model depends on two physical
parameters: mass $M$ of the elementary exitation and
$$
\kappa={\pi\beta^2\over8\pi+\beta^2}.
\eqno(5.2)
$$
The $S$-matrix of the boson $\varphi(x)$ is given by$^{10}$
$$
S(\theta)={\tanh\half(\theta-i\kappa)\over\tanh\half(\theta+i\kappa)}
={\sinh\theta-i\sin\kappa\over\sinh\theta+i\sin\kappa}.
\eqno(5.3)
$$
The bosonic sector of the sinh-Gordon model has been
well investigated.$^{11}$ Every form factor
$F_\phi^{(n)}(\theta_1,\cdots,\theta_n)$,
$\phi\in[\tau]\oplus[\varphi]$ takes the form
$$
F_\phi^{(n)}(\theta_1,\cdots,\theta_n)
=P_\phi^{(n)}(e^{\theta_1},\cdots,e^{\theta_n})
\prod_{i<j}{\tilde{F}(\theta_i-\theta_j)\over e^{\theta_i}+e^{\theta_j}},
\eqno(5.4)
$$
where $P_\phi^{(n)}(x_1,\cdots,x_n)$ is a symmetric polynomial,
$\tilde{F}(\theta)$ is so called minimal two-point form factor
$$
\eqalign{
&\tilde{F}(\theta)=\tilde{F}(-\theta)S(\theta),
\cr
&\tilde{F}(\theta)=\tilde{F}(2\pi i-\theta),
\cr
&\lim_{\theta\rightarrow\infty}\tilde{F}(\theta)=1,
}\eqno(5.5)
$$
and $\tilde{F}(\theta)$ has no poles in the strip $0<\hbox{ Im }\theta<\pi$.
Explicit form of the minimal form factor for the sinh-Gordon model is$^{11}$
$$
\eqalign{
\tilde{F}(\theta)
&={\cal N}\exp\left[8\int_0^\infty{dx\over x}
{\sinh{\kappa x\over2\pi}\,\sinh\half\left(1-{\kappa\over\pi}\right)x
\,\sinh{x\over2}\over\sinh^2x}\sinh^2{(i\pi-\theta)x\over2\pi}\right],
\cr
{\cal N}
&=\exp\left[-4\int_0^\infty{dx\over x}
{\sinh{\kappa x\over2\pi}\,\sinh\half\left(1-{\kappa\over\pi}\right)x
\,\sinh{x\over2}\over\sinh^2x}\right].
}\eqno(5.6)
$$
There is an additional property of $\tilde{F}(\theta)$ specific for the
sinh-Gordon model
$$
\tilde{F}(\theta+i\pi)\tilde{F}(\theta)
={\sinh\theta\over\sinh\theta+i\sin\kappa}.
\eqno(5.7)
$$
Polynomials $P_\phi^{(n)}(x_1,\cdots,x_n)$ satisfy the equation
$$
\eqalign{
&(-)^nP_\phi^{(n+2)}(-x,x,x_1,\cdots,x_n)
=xD_n^-(x,x_1,\cdots,x_n)P_\phi^{(n)}(x_1,\cdots,x_n).
\cr
&D_n^-={-i\over\tilde{F}(i\pi)}
\left[\prod_{i=1}^n(x+e^{i\kappa}x_i)(x-e^{-i\kappa}x_i)
-\prod_{i=1}^n(x-e^{i\kappa}x_i)(x+e^{-i\kappa}x_i)\right].
}\eqno(5.8)
$$
It follows from Eq. (3.4).

For the basic field $\varphi(x)$ and the trace  of the energy-momentum
tensor $\tau(x)=T_\mu^\mu(x)=T_{+-}(x)$ the polynomials take the form$^{11}$
$$
\eqalign{
&P_\varphi^{(1)}(x)=1,
\cr
&P_\varphi^{(2n+1)}(x_1,\cdots,x_{2n+1})
=\left({4\sin\kappa\over\tilde{F}(i\pi)}\right)^n
\sigma_{2n+1}^{(2n+1)}p_{2n+1}(x_1,\cdots,x_{2n+1}),\quad n\geq1,
}\eqno(5.9{\rm a})
$$
$$
\eqalign{
&P_\tau^{(2)}(x_1,x_2)={M^2\over\tilde{F}(i\pi)}(x_1+x_2),
\cr
&P_\tau^{(2n)}(x_1,\cdots,x_{2n})
={M^2\over\tilde{F}(i\pi)}\left({4\sin\kappa\over\tilde{F}(i\pi)}\right)^{n-1}
\sigma_1^{(2n)}\sigma_{2n-1}^{(2n)}p_{2n}(x_1,\cdots,x_{2n}),
\quad n\geq2,
}\eqno(5.9{\rm b})
$$
where $\sigma_i^{(n)}=\sigma_i^{(n)}(x_1,\cdots,x_n)$ are elementary
symmetric polynomials
$$
\eqalign{
&\sigma_1^{(n)}=x_1+x_2+\cdots+x_n,
\cr
&\sigma_2^{(n)}=x_1x_2+x_1x_3+\cdots+x_{n-1}x_n,
\cr
&\hbox to 25 em{\leaders\hbox{$\cdot$\hskip 1 em}\hfil}
\cr
&\sigma_n^{(n)}=x_1x_2\cdots x_n.
}$$
It is easy to cheque using Eq. (5.9b) that the Hamiltonian from
Eq. (2.7) can be performed as
$$
{\cal H}=\int dx^1T_0^0(x).
\eqno(5.10)
$$
Polynomials $p_n(x_1,\cdots,x_n)$ for several first $n$'s are$^{11}$
$$
\eqalign{
p_3(x_1,\cdots,x_3)
&=1,
\cr
p_4(x_1,\cdots,x_4)
&=\sigma_2,
\cr
p_5(x_1,\cdots,x_5)
&=\sigma_2\sigma_3-4\cos^2\kappa\cdot\sigma_5,
\cr
p_6(x_1,\cdots,x_6)
&=\sigma_2\sigma_3\sigma_4
-4\cos^2\kappa\cdot(\sigma_4\sigma_5+\sigma_1\sigma_2\sigma_6)
-(1-4\cos^2\kappa)\sigma_3\sigma_6,
\cr
p_7(x_1,\cdots,x_7)
&=\sigma_2\sigma_3\sigma_4\sigma_5
-4\cos^2\kappa\cdot(\sigma_4\sigma_5^2+\sigma_1\sigma_2\sigma_6
+\sigma_2^2\sigma_3\sigma_7)
\cr
&+16\cos^4\kappa\cdot\sigma_2\sigma_5\sigma_7
-(1-4\cos^2\kappa)(\sigma_1\sigma_2\sigma_4\sigma_7+\sigma_3\sigma_5\sigma_6)
\cr
&-(1-4\cos^2\kappa)^2\sigma_1\sigma_6\sigma_7
-4\cos^2\kappa\cdot(1-4\cos^2\kappa)\sigma_7^2,
}\eqno(5.11)
$$
where obvious superscripts of $\sigma_i^{(n)}$ are omitted.

Consider now fields with $\zeta_\phi=-1$. We shall search form
factors in the form
$$
F_\phi^{(n)}(\theta_1,\cdots,\theta_n)
=e^{\half\sum\theta_i}Q_\phi^{(n)}
\left(e^{\theta_1},\cdots,e^{\theta_n}\right)
\prod_{i<j}{\tilde{F}(\theta_i-\theta_j)\over e^{\theta_i}+e^{\theta_j}},
\eqno(5.12)
$$
where $Q_\phi^{(n)}(x_1,\cdots,x_n)$ is a rational symmetric function
that has no poles out of zero. The first multiplier in the right
hand side provides $\lambda_\phi=-1$ in Eq. (3.3). Eq. (3.4) leads to
the equation
$$
\eqalign{
&(-)^nQ_\phi^{(n+2)}(-x,x,x_1,\cdots,x_n)
=D_n^+(x,x_1,\cdots,x_n)Q_\phi^{(n)}(x_1,\cdots,x_n).
\cr
&D_n^+={-1\over\tilde{F}(i\pi)}
\left[\prod_{i=1}^n(x+e^{i\kappa}x_i)(x-e^{-i\kappa}x_i)
+\prod_{i=1}^n(x-e^{i\kappa}x_i)(x+e^{-i\kappa}x_i)\right].
}\eqno(5.13)
$$

Consider a spinor field $\psi(x)=(\psi_+(x),\psi_-(x))$ with the Lorentz
spin $s_{\psi_\pm}=\pm\half$. It is convenient to slightly modify
Eq. (5.12). Namely, let
$$
\eqalign{
&F_{\psi_\pm}^{(2n+1)}(\theta_1,\cdots,\theta_{2n+1})
\cr
&\qquad=\sqrt{M}e^{\mp\half({i\pi\over2}+\sum\theta_i)}Q^{(2n+1)}(e^{\mp\theta_1},
\cdots,e^{\mp\theta_{2n+1}})
\prod_{i<j}{\tilde{F}(\theta_i-\theta_j)\over
e^{\mp\theta_i}+e^{\mp\theta_j}},
}\eqno(5.14)
$$
The total degree of $Q^{(2n+1)}(x_1,\cdots,x_n)$ is determined by the spin
and is equal to $2n^2$. Using Eq. (2.16) we get that $\psi(x)$
is a Majorana fermion
$$
\psi_\alpha^+(x)=\psi_\alpha(x).
\eqno(5.15)
$$

Suppose that the form factors of $\psi_\pm$
behave as $\exp(\pm\half\theta_i)$ as $\theta_i\longrightarrow\pm\infty$
(as for a free fermion). Then $Q^{(2n+1)}$ must be a polynomial. The
last condition determines the first few $Q$'s uniquely:
$$
\eqalign{
Q^{(1)}(x)
&=1,
\cr
Q^{(3)}(x_1,\cdots,x_3)
&=\left(-{2\over\tilde{F}(i\pi)}\right)(\sigma_1^2+\sigma_2),
\cr
Q^{(5)}(x_1,\cdots,x_5)
&=\left(-{2\over\tilde{F}(i\pi)}\right)^2
[(\sigma_1^2+\sigma_2)(\sigma_3^2+\sigma_2\sigma_4)
\cr
&+(1-4\cos^2\kappa)\sigma_1\sigma_3\sigma_4
-2(1+\cos^2\kappa)\sigma_4^2-4\cos^2\kappa\cdot\sigma_1^3\sigma_5
\cr
&-\sigma_1\sigma_2\sigma_5-\sigma_3\sigma_5].
}\eqno(5.16)
$$
This Majorana fermion has the pair $S$-matrix
$$
S_\psi(\theta)=-{\tanh\half(\theta-i\kappa)\over\tanh\half(\theta+i\kappa)}.
\eqno(5.17)
$$

Similarly we can find the form factors of a field from the `disorder'
sector. We shall search for a scalar field, $\mu(x)$, whose form factors
are finite at the infinity. Let
$$
F_{\mu}^{(2n)}(\theta_1,\cdots,\theta_{2n})
=e^{\pm\half\sum\theta_i}Q^{(2n)}(e^{\pm\theta_1},
\cdots,e^{\pm\theta_{2n}})
\prod_{i<j}{\tilde{F}(\theta_i-\theta_j)\over
e^{\pm\theta_i}+e^{\pm\theta_j}}
\eqno(5.18)
$$
with some polynomials $Q^{(2n)}(x_1,\cdots,x_{2n})$ of the total degree
$2n(n-1)$. First polinomials are
$$
\eqalign{
Q^{(0)}
&=1,
\cr
Q^{(2)}(x_1,x_2)
&=\left(-{2\over\tilde{F}(i\pi)}\right),
\cr
Q^{(4)}(x_1,\cdots,x_4)
&=\left(-{2\over\tilde{F}(i\pi)}\right)^2
(-4\cos^2\kappa\cdot\sigma_4+\sigma_1\sigma_3+\sigma_2^2).
}\eqno(5.19)
$$
We see that at least for the first form factors the choice of the
sign in Eq. (5.18) is not essential.
\medskip\noindent
{\bf 6. Conclusion}

\nobreak\medskip\nobreak\noindent
We see that at least in one-particle models with ${\bf Z_2}$ symmetry
there are three sectors of mutually local fields: bosonic, fermionic
and `disorder' ones. Fields from different sectors are generally mutually
nonlocal. The bosonic sector contains one neutral boson with the
$S$-matrix $S(\theta)$. The fermionic sector contains one
Majorana fermion with the $S$-matrix $-S(\theta)$. We stress that
for constructing a full basis in the state space it is enough to
act on the vacuum by either bosons or fermions. Both bosons and
fermions discribe the same dynamical system. They create the same
asymptotic states. The `disorder' fields do not create any asymptotic
states. The role of `disorder' fields seems somewhat mysterious.

The situation is different in the theories without ${\bf Z_2}$ symmetry.
For example, the reduced sin-Gordon model with the $S$-matrix
$$
S(\theta)={\tanh\half\left(\theta+{2\pi i\over3}\right)
\over\tanh\half\left(\theta-{2\pi i\over3}\right)}
$$
contains only one particle $b$ and allows the virtual fusion
$b+b\longrightarrow b$. This fusion imposes a connection
between even and odd form factors of the particle $b$.$^{12}$
It means that fermions are forbidden. But there are at least two
bosonic sectors in these theories. Particle fields from different
sectors are mutually nonlocal.$^{13}$

In Sec. 5 we discussed the sinh-Gordon theory. We cited several
leading form factors in each sector. But for this model explicit integral
formulas for form factors in the bosonic sector are known.$^{13}$
It would be important to construct integral representations for
form factors in other sectors. It would proove the solubility
of the form factor axioms in fermionic and `disorder' sectors.

We deformed the form factor axioms very slightly introducing
one sign factor. It would be interesting to find more essential
deformations preserving locality of fields.
\medskip\noindent
{\bf Acknoledgments}

\nobreak\medskip\nobreak\noindent
The author is grateful to A.~V.~Antonov, A.~A.~Belov, A.~A.~Kadeishvili,
S.~E.~Par\-kho\-men\-ko and V.~V.~Postnikov for useful discussions.
\medskip\noindent
{\bf References}

\nobreak\medskip\nobreak
\item{1.}E.~C.~Marino, B.~Schroer, J.~A.~Swieca,
{\it Nucl. Phys.} {\bf B200 [FS4]}, 473 (1982)

\item{2.}Vl.~S.~Dotsenko, A.~M.~Polyakov,
{\it Adv. Stud. Pure Math.} {\bf 16}, 171 (1988)

\item{3.}M.~Sato, M.~Jimbo, T.~Miwa, {\it Publ. RIMS} {\bf 14}, 223 (1978);
{\bf 15}, 577; 871 (1979); {\bf 16}, 531 (1980)

\item{4.}S.~Colemann, {\it Phys. Rev.} {\bf D11}, 2088 (1975)

\item{5.}T.~R.~Klassen, E.~Melzer, {\it Int. J. Mod. Phys.} {\bf A8},
4131 (1993)

\item{6.}A.~B.~Zamolodcikov, Al.~B.~Zamolodchikov, {\it Annals Phys.}
{\bf 120}, 253 (1979)

\item{7.}M.~Karowski, P.~Weisz, {\it Nucl. Phys.} {\bf B139}, 455 (1978)

\item{8.}F.~A.~Smirnov, {\it in} Introduction to quantum group
and integrable massive models of quantum field theory, Nakai
Lectures on Mathematical Physics (World Scientific, Singapore 1990)

\item{9.}B.~Berg, M.~Karowski, P.~Weisz, {\it Phys. Rev.} {\bf D19},
2477, 1979

\item{10.}A.~E.~Arinshtein, V.~A.~Fateev, A.~A.~Zamolodchikov,
Phys. Lett. {\bf B87}, 389 (1979)

\item{11.}A.~Fring, G.~Mussardo, P.~Simonetti, {\it Nucl. Phys.}
{\bf B393}, 413 (1993)

\item{12.}F.~A.~Smirnov, {\it Int. J. Mod. Phys.} {\bf A4}, 4213 (1989)

\item{13.}F.~A.~Smirnov, {\it Nucl. Phys.} {\bf B337}, 156 (1990)
\bye